\documentclass[12pt]{article}

\usepackage{amsmath}
\usepackage{newtxtext}
\usepackage[subscriptcorrection]{newtxmath}
\usepackage{bm}
\usepackage{comment}
\usepackage{graphicx}
\usepackage[plain,noend]{algorithm2e}
\usepackage{apacite}  
\usepackage{geometry} 
\usepackage{natbib}
\newtheorem{theorem}{Theorem}
\usepackage{authblk}

\usepackage{multirow}
\newtheorem{proof}{Proof}

\makeatletter
\renewcommand{\algocf@captiontext}[2]{#1\algocf@typo. \AlCapFnt{}#2} 
\def\@algocf@capt@plain{top}
\renewcommand{\algocf@makecaption}[2]{%
  \addtolength{\hsize}{\algomargin}%
  \sbox\@tempboxa{\algocf@captiontext{#1}{#2}}%
  \ifdim\wd\@tempboxa >\hsize
    \hskip .5\algomargin%
    \parbox[t]{\hsize}{\algocf@captiontext{#1}{#2}}
  \else%
    \global\@minipagefalse%
    \hbox to\hsize{\box\@tempboxa}
  \fi%
  \addtolength{\hsize}{-\algomargin}%
}
\makeatother

\begin{document}

\title{\bf Association measures for two-way contingency tables based on multi-categorical proportional reduction in error}

\author[1]{Wataru Urasaki}
\author[1]{Kouji Tahata}
\author[2]{Sadao Tomizawa}

\affil[1]{Department of Information Science and Technology, Tokyo University of Science}
\affil[2]{Department of Information Sciences, Tokyo University of Science}
\date{}

\maketitle

\begin{abstract}
In two-way contingency tables under an asymmetric situation, where the row and column variables are defined as explanatory and response variables, respectively, quantifying the extent to which the explanatory variable contributes to predicting the response variable is important. 
One quantification method is the association measure, which indicates the degree of association in a range from $0$ to $1$.
Among various measures that have been proposed, those based on proportional reduction in error (PRE) are particularly notable for their simplicity and intuitive interpretation. 
These measures, including Goodman–Kruskal's lambda proposed in 1954, are widely implemented in statistical software such as R and SAS and remain extensively used. 
However, a well-known limitation of PRE measures is their potential to return a value of $0$ despite no independence. 
This issue arises because the measures are constructed based solely on the maximum joint and marginal probabilities, failing to make full use of the information available in the contingency table.
To address this problem, we propose an extension of PRE measures designed for the proportional reduction in error with multiple categories. 
The properties of the proposed measures are examined, and their utility is demonstrated through numerical experiments. 
The results suggest their potential as practical tools in applied statistics.
\end{abstract}

\medskip


\medskip


\section{Introduction}\label{sec1}
Categorical variables are defined by discrete categories and are used across a wide range of fields such as medicine, psychology, and social sciences.
We consider two categorical variables, $X$ and $Y$, where $X$ consists of $r$ categories and $Y$ consists of $c$ categories.
These two variables yield $r \times c$ combinations, which can be represented in a table with $r$ rows and $c$ columns.
Such a table is called a two-way contingency table, in which each ($i, j$) cell ($i=1,\ldots, r;j=1,\ldots, c$) displays the observed frequencies.
The primary objective of contingency table analysis is to evaluate the independence between variables, for which methods such as Pearson's chi-square test and Fisher’s exact test are widely used.
Furthermore, extending the concept of independence, research aimed at elucidating relationships between variables has a long history and has been extensively explored within the framework of association models.
Various association models have been proposed by \cite{goodman1981association, goodman1985analysis}, \cite{rom1992generalized}, and \cite{kateri2018phi}.
In parallel, studies on association measures have sought to quantify the strength of association between variables through various approaches.
Several measures, such as Cram\'{e}r's coefficient (\citealp{cramer1946mathematical}), Theil's uncertainty coefficient (\citealp{theil1970estimation}), and Goodman–Kruskal's lambda and tau (\citealp{doi:10.1080/01621459.1954.10501231}), remain widely used today. 
These measures are implemented in statistical software packages such as R and SAS, making them readily available to researchers and practitioners.
In recent years, \cite{tomizawa2004generalization}, \cite{urasaki2023generalized} have proposed association measures based on generalized divergence, and \cite{kvaalseth2017alternative, kvaalseth2023association} reconsidered association measures by focusing on desirable properties, particularly value-validity.

Among the various association measures that have been proposed, some are categorized as asymmetric measures, which are designed for situations in which the roles of the row and column variables are predetermined as explanatory and response variables, respectively. 
Goodman–Kruskal's lambda, denoted by $\lambda_{Y\mid X}$, is a representative asymmetric measure based on the concept of proportional reduction in error (PRE). 
Because of its simple construction and intuitive interpretation, this measure has been widely applied across a variety of fields (e.g., \cite{jaroszewicz2004goodman, suich2003asymptotic, tichy2011evaluating}). 
However, existing PRE measures have a well-known limitation: they are constructed using only the category with the largest probability, namely the maximum marginal probability and, within each row, the maximum joint probability. 
Consequently, they do not make full use of the information contained in the contingency table, and may take the value $0$ even when the variables are not independent. 
To overcome this limitation, $\lambda^K_{Y\mid X}$ was proposed by \cite{kvaalseth2018measuring} as an alternative to $\lambda_{Y\mid X}$. 
Although $\lambda^K_{Y\mid X}$ improves the value-validity property by ensuring that independence is indicated when its value is $0$, it may still fail to adequately reflect the degree of association in situations where relevant information is distributed across more than one category.

From the perspective of interpretability, PRE-based measures remain particularly attractive. 
As argued by \cite{costner1965criteria}, measures of association should be operationally interpretable in terms of the proportional reduction in prediction error achieved by using information on the explanatory variable. 
This viewpoint provides a useful principle for constructing and selecting association measures, since it gives them a clear practical meaning in terms of predictive improvement. 
Motivated by this perspective and by the limitations of existing PRE measures, this paper proposes a multi-categorical extension of PRE measures. 
The proposed measures incorporate information from multiple categories rather than relying solely on the single most probable category, thereby providing a more informative assessment of association while retaining the simple and meaningful PRE interpretation.

In this paper, we investigate the construction and properties of the proposed measures and clarify their relationships with $\lambda_{Y\mid X}$ and $\lambda^K_{Y\mid X}$. 
We also present numerical experiments to illustrate situations in which existing PRE measures do not perform satisfactorily and to demonstrate the practical usefulness of the proposed measures. 
Through these analyses, we show that the proposed multi-categorical PRE measures offer a flexible and interpretable framework for evaluating association in contingency tables, regardless of whether the information relevant to prediction is concentrated in a single category or distributed across multiple categories.

\section{Association measures based on proportional reduction in error}\label{sec2}
Consider an $r \times c$ contingency table in which the row variable $X$ and the column variable $Y$ have nominal categories, and assume the asymmetric setting in which $X$ is the explanatory variable and $Y$ is the response variable.
Let $p_{ij}$ denote the probability that an observation falls in the $i$th row and $j$th column of the table ($i=1,\ldots, r;j=1,\ldots, c)$.
In addition, the marginal probabilities are defined as $p_{i+}=\sum_{j=1}^c p_{ij}$ and $p_{+j}=\sum_{i=1}^r p_{ij}$, respectively.
In this section, we review the association measures that have been proposed based on the concept of proportional reduction in error (PRE) and discuss their properties and applications.

\subsection{Goodman–Kruskal’s lambda}\label{sec2-1}
Association measures based on proportional reduction in error were proposed to quantify the degree to which the probability of error of the response variable $Y$ in Case 1 (when no information is available) differs from that in Case 2 (when information on the explanatory variable $X$ is provided).
Therefore, in Case 1, the best prediction for $Y$ is obtained by selecting the category of $Y$ with the highest marginal probability, giving a probability of error equal to $1-\underset{j}{\max} \{ p_{+j} \}$. 
In Case 2, the best prediction is based on the category of $Y$ with the highest conditional probability given the observed category of $X$, giving a probability of error equal to $1-\sum_{i=1}^r \underset{j}{\max} \{ p_{ij} \}$, which equals the weighted average of the within-row conditional error probabilities $\sum^r_{i=1}p_{i+}(1-\underset{j}{\max} \{ p_{ij} \}/p_{i+})$. 
Based on these error probabilities, \cite{doi:10.1080/01621459.1954.10501231} proposed the following PRE measure $\lambda_{Y\vert X}$:
\begin{align*}
\lambda_{Y\vert X} &= \frac{(\text{Prob. of error in Case 1}) - (\text{Prob. of error in Case 2})}{(\text{Prob. of error in Case 1})} \\
&= \frac{\sum_{i=1}^r p_{im_i}- p_{+ m_0}}{1-p_{+ m_0}},
\end{align*}
where
\begin{align*}
p_{+m_0} = \underset{j}{\max} \{ p_{+j} \}, \hspace{0.3cm}
p_{im_i} = \underset{j}{\max} \{ p_{ij} \}.
\end{align*}
This measure represents the relative reduction in prediction error for $Y$ when $X$ is known compared to when $X$ is unknown.
In other words, $\lambda_{Y\vert X}$ represents the proportion of errors that can be eliminated by taking into account the knowledge of $X$.
The asymmetric measure $\lambda_{Y\vert X}$ possesses several important properties: its values range between 0 and 1; it equals $0$ when $X$ and $Y$ are independent, and it equals $1$ when each row of the contingency table contains at most one positive cell probability $p_{ij}>0$.
Goodman–Kruskal's lambda is implemented in several major statistical software packages, including SAS, SPSS, R, and others.

Although asymmetric measures provide simple and intuitively appealing definitions and interpretations, they have a well-known limitation.
For $\lambda_{Y \vert X}$, the measure can equal $0$ whenever all $p_{im_i}$ for $i=1,\ldots, r$ occur in the same column of the contingency table, even though $X$ and $Y$ may not be statistically independent.
As several authors have noted, this problem tends to arise when the marginal probabilities are far from uniform (e.g. \citealp{everitt1992analysis}, and \citealp{reynolds1977analysis}), and the use of these measures may be inappropriate when the marginal distributions are highly skewed. 
To address this issue, \cite{stavig1979alternatives} discussed other measures that could be used instead, while \cite{kvaalseth2018measuring} proposed an alternative measure based on a simple modification of Goodman–Kruskal's lambda.

\subsection{Alternative measures to the Goodman–Kruskal's lambda}\label{sec2-2}
\cite{kvaalseth2018measuring} proposed the following alternative measure to address the limitations of Goodman–Kruskal's $\lambda_{Y \vert X}$:
\begin{align*}
\lambda^{K}_{Y\vert X} &= \frac{\sqrt{\sum^r_{i=1} p_{i m_{i}}^2 /p_{i+} } - p_{+ m_{0}} }{1-p_{+ m_{0}}}.
\end{align*}
This measure replaces the probability of error in Case 2 of Goodman–Kruskal's $\lambda_{Y \vert X}$ with 
\begin{align*}
(\text{Prob. of error in Case 2}) &= 1-\left\{\sum^r_{i=1}p_{i+}\left(\frac{p_{i m_{i}}}{p_{i+}}\right)^2\right\}^{1/2}.
\end{align*}
In the above formula, the probability of error in Case 2 is calculated by applying the weighted quadratic mean or the weighted root mean square (RMS), which is commonly used in statistics and other fields such as engineering and physics applications, to the conditional probabilities. 
Although the interpretation of the error probability differs slightly from that in $\lambda_{Y \vert X}$, it remains sufficiently simple and meaningful for measuring association via proportional reduction in error.

The alternative measure $\lambda^{K}_{Y\vert X}$ retains an important property of an association measure: $\lambda^{K}_{Y\vert X}=0$ if and only if $X$ and $Y$ are independent, in the sense that $p_{ij}=p_{i\cdot}p_{\cdot j}$ holds for $i = 1, \ldots, r$ and at least for the column $j$ corresponding to $p_{+ m_{0}}$.
In particular, when $c = 2$ and $\lambda^{K}_{Y\vert X}=0$, this indicates a completely independent structure between the variables $X$ and $Y$.
Moreover, consider a scenario in which $\lambda_{Y\vert X}=0$ despite the absence of independence.
This occurs when the maximum joint probability in each row is concentrated within a single column, so that $p_{+m_0} = \sum^r_{i=1}p_{im_0}$.
In such cases, $\lambda^K_{Y\vert X}\neq 0$, enabling the detection and quantification of association.
In addition, $\lambda^K_{Y\vert X}$ also possesses the value-validity property held by $\lambda_{Y\vert X}$ (for details, see \citealp{kvaalseth2017alternative}), along with other practical advantages. 

\section{Association measures based on multi-categorical proportional reduction in error}\label{sec3}
In the previous section, we introduced Goodman–Kruskal's $\lambda_{Y\vert X}$, a representative association measure based on proportional reduction in error, along with its alternative measure $\lambda^K_{Y\vert X}$.
While these methods are important for measuring association, they have a limitation that restricts their practical applicability.
Specifically, when the maximum joint probabilities in each row are concentrated in a single column and exhibit partial independence, both $\lambda_{Y\vert X}$ and $\lambda^K_{Y\vert X}$ become $0$, rendering the measures unable to detect associations that may exist in other columns.
This issue arises because both measures rely solely on the category with the highest probability when calculating the probabilities of error, which means much of the information observed in the contingency table is discarded rather than utilized.
Therefore, in this section, we focus on proportional reduction in error across multiple categories and propose an extended association measure. 
This measure is easily interpretable and well-suited for exploratory applied research.

\subsection{Goodman–Kruskal's lambda based on multi-categorical proportional reduction in error}
Consider an asymmetric situation where the row variable $X$ serves as the explanatory variable and the column variable $Y$ as the response variable.
Then we propose the following association measure based on multi-categorical proportional reduction in error:
For $t$ given an integer in $1 \leq t < c$,
\begin{align*}
\lambda^{(t)}_{Y\vert X} &= \frac{\sum^r_{i=1} \sum^t_{k=1}p_{i m_{i(k)}}- \sum^t_{k=1}p_{+ m_{0(k)}} }{1-\sum^t_{k=1}p_{+ m_{0(k)}}},
\end{align*}
where $p_{+ m_{0(k)}}$ and $p_{i m_{i(k)}}$ are the $k$-th largest value in $\{p_{+1}, \ldots, p_{+c}\}$ and $\{p_{i1}, \ldots, p_{ic}\}$, respectively.
We assume that $\sum^t_{k=1}p_{+ m_{0(k)}} \neq 1$.
Additionally, when $t=1$, $\lambda^{(1)}_{Y\vert X}$ coincides with Goodman–Kruskal's $\lambda_{Y\vert X}$.
The measure $\lambda^{(t)}_{Y\vert X}$ defines the probabilities of error in Case 1 and Case 2 as follows:
\begin{align*}
(\text{Prob. of error in Case 1}) &= 1-\sum^t_{k=1}p_{+ m_{0(k)}}, \\
(\text{Prob. of error in Case 2}) &= 1-\sum^r_{i=1} \sum^t_{k=1}p_{i m_{i(k)}}.
\end{align*}
By construction, the interpretation of $\lambda^{(t)}_{Y\vert X}$ remains as simple as that of $\lambda_{Y\vert X}$, while the interpretation of probabilities of error is extended to cover multiple categories.
Specifically, $\lambda_{Y\vert X}$ considered only the maximum joint and marginal probabilities to be critical for measuring association. 
In contrast, this approach emphasizes that information from categories with higher probabilities is also important for estimating association.

Some important properties of $\lambda^{(t)}_{Y\vert X}$ are as follows:
\begin{theorem}\label{t1}
For fixed $t$, the measure $\lambda^{(t)}_{Y\vert X}$ satisfies the following properties:
\begin{itemize}
\item[1-1] $0 \leq \lambda^{(t)}_{Y\vert X} \leq 1$.
\item[1-2] If $p_{im_{i(k)}}=p_{i+}p_{+m_{0(k)}}$ ($1 \leq k \leq t$) holds, then $\lambda^{(t)}_{Y\vert X}=0$ (also if the independence $\{p_{ij}=p_{i+}p_{+j}\}$ holds), but the converse need not hold.
\item[1-3] $\lambda^{(t)}_{Y\vert X}=1$ if and only if $p_{i+}=\sum^t_{k=1}p_{im_{i(k)}}$ for $i=1, \dots, r$.
\end{itemize}
\end{theorem}
Theorem \ref{t1}-3 means that the response variable $Y$ of a randomly chosen observation is perfectly predictable from the explanatory variable $X$, that is, if each row of the contingency table contains at most $t$ nonzero cell probabilities $p_{ij}$.
For proof of Theorem \ref{t1}, see the Appendix \ref{secA1}. 
The advantage of this measure lies in its ability to address the issues described above simply by calculating the measure for $t \geq 2$.
For instance, under the condition with $p_{+m_{0(1)}} = \sum^r_{i=1}p_{im_{0(1)}}$, the proposed measure with $t = 2$ is given as follows:
\begin{align*}
\lambda^{(2)}_{Y\vert X} &= \frac{\sum^2_{k=1}\left(\sum^r_{i=1}p_{i m_{i(k)}}- p_{+ m_{0(k)}} \right)}{1-\sum^2_{k=1}p_{+ m_{0(k)}}} 
= \frac{\sum^r_{i=1}p_{i m_{i(2)}}- p_{+ m_{0(2)}} }{1-\sum^2_{k=1}p_{+ m_{0(k)}}}.
\end{align*}
$\lambda^{(2)}_{Y\vert X}$ measures the degree of association based on the probabilities of error calculated not only from the maximum marginal probability but also from the second-largest observed joint and marginal probabilities.
As a result, it avoids the limitations caused by the $j$-th column corresponding to $p_{+m_{0(1)}}$, without affecting the overall or marginal probabilities.
Moreover, calculating $\lambda^{(t)}_{Y\vert X}$ for all $1 \leq t < c$ can provide valuable insights in exploratory research.
Goodman–Kruskal's $\lambda_{Y\vert X}$, which relies only on the most frequent category, can be interpreted as a measure quantifying how much interpretability $X$ has for predicting the most frequent category of the response variable $Y$.
However, in exploratory research, predicting only the most frequent category of $Y$ may not suffice.
In some cases, it is essential to understand the extent to which multiple high-frequency categories of $Y$ can be explained by $X$.
Therefore, when the purpose of the survey is to quantify interpretability across multiple categories or to compare different contingency tables under such situations, our proposed measure is particularly well-suited.
However, this proposal does not fully address the limitations regarding independence. 
To resolve the issue, an alternative measure is also presented in the following subsection.

\subsection{Alternative measure based on multi-categorical proportional reduction in error}
We propose an alternative measure based on multi-categorical proportional reduction in error, defined as follows:
For $t$ given an integer in $1 \leq t < c$,
\begin{align*}
\lambda^{K(t)}_{Y\vert X} &= \frac{\sqrt{\sum^r_{i=1} \left(\sum^t_{k=1}p_{i m_{i(k)}} \right)^2 /p_{i+} } - \sum^t_{k=1}p_{+ m_{0(k)}} }{1-\sum^t_{k=1}p_{+ m_{0(k)}}}.
\end{align*}
We also assume that $\sum^t_{k=1}p_{+ m_{0(k)}} \neq 1$.
Additionally, when $t=1$, $\lambda^{K(1)}_{Y\vert X}$ coincides with the alternative measure $\lambda^K_{Y\vert X}$.
The measure $\lambda^{K(t)}_{Y\vert X}$ defines the probabilities of error in Case 2 as follows:
\begin{align*}
(\text{Prob. of error in Case 2}) &= 1-\left\{\sum^r_{i=1} \sum^t_{k=1}p_{i+}\left(\frac{p_{i m_{i(k)}}}{p_{i+}}\right)^2\right\}^{1/2}.
\end{align*}
The measure $\lambda^{K(t)}_{Y\vert X}$ also has the same simple and meaningful interpretation as a PRE measure: it shows the proportional reduction in prediction error probability afforded by specifying the explanatory variable $X$.
Moreover, since 
\begin{align*}
\sum^r_{i=1} \sum^t_{k=1}p_{i+}\left(\frac{p_{i m_{i(k)}}}{p_{i+}}\right)^2 \geq \left\{ \sum^r_{i=1} \sum^t_{k=1}p_{i+}\left(\frac{p_{i m_{i(k)}}}{p_{i+}}\right) \right\}^2
\end{align*}
holds by Jensen's inequality, $\lambda^{K(t)}_{Y\vert X} \geq \lambda^{(t)}_{Y\vert X}$ always follows for any $t$.

Some important properties of $\lambda^{K(t)}_{Y\vert X}$ are as follows:
\begin{theorem}\label{t2}
For fixed $t$, the measure $\lambda^{K(t)}_{Y\vert X}$ satisfies the following properties:
\begin{itemize}
\item[2-1] The value of $\lambda^{K(t)}_{Y\vert X}$ is between $0$ and $1$ inclusive.
\item[2-2] 
If $\lambda^{K(1)}_{Y\vert X} = \cdots = \lambda^{K(t-1)}_{Y\vert X}=0$ and $p_{im_{i(k)}}=p_{i+}p_{+m_{0(k)}}$ ($1 \leq k \leq t$), then $\lambda^{K(t)}_{Y\vert X}=0$, (also if the independence $\{p_{ij}=p_{i+}p_{+j}\}$ holds).
\item[2-3] $\lambda^{K(t)}_{Y\vert X}=1$ if and only if $p_{i+}=\sum^t_{k=1}p_{im_{i(k)}}$ for $i=1, \dots, r$.
\end{itemize}
\end{theorem}
Theorem \ref{t2}-2 shows that if $\lambda^{K(1)}_{Y\vert X} = \cdots = \lambda^{K(t)}_{Y\vert X}=0$ for all $t=1,\dots, c-1$, then there is complete independence in the contingency table.
For a proof of Theorem \ref{t2}, see the Appendix \ref{secA2}. 
This measure is an extension of $\lambda^K_{Y\vert X}$, based on a similar concept to the proposal for $\lambda^{(t)}_{Y\vert X}$.
As a result, it shares similar advantages with $\lambda^K_{Y\vert X}$.
However, the primary benefit of the extension is that it completely resolves the limitations regarding independence, which $\lambda^{K}_{Y\vert X}$ could not address.
The usage of $\lambda^{K(t)}_{Y\vert X}$ is also well-suited for comprehensively investigating data by calculating its values for all $1 \leq t < c$.
This approach enables a more thorough exploration of the data across multiple categories.

\subsection{Comments on symmetric situation}
\cite{doi:10.1080/01621459.1954.10501231} also proposed a PRE measure for the symmetric situation where it is unnecessary to distinguish between $X$ and $Y$ as explanatory and response variables, as follows:
\begin{align*}
\lambda &= w_y \lambda_{Y\vert X} + w_x \lambda_{X\vert Y} 
= \frac{\sum_{i=1}^r p_{im_i} + \sum_{j=1}^c p_{M_j j} - p_{+ m_0} - p_{M_0+}}{2-p_{+m_0}+p_{M_0+}},
\end{align*}
where 
\begin{align*}
w_{y} &= \frac{1-p_{+ m_{0}}}{2-p_{+m_0}+p_{M_0+}}, \quad
w_{x} = \frac{1-p_{M_{0}+}}{2-p_{+m_0}+p_{M_0+}}.
\end{align*}
The symmetric measure $\lambda$ is defined as a weighted mean of $\lambda_{Y\vert X}$ and $\lambda_{X\vert Y}$ (which reverses the roles of $X$ and $Y$) given by
\begin{align*}
\lambda_{X\vert Y} &= \frac{\sum_{j=1}^c p_{M_j j}- p_{M_0+}}{1-p_{M_0+}},
\end{align*}
where $p_{M_0+} = \underset{i}{\max} \{ p_{i+} \}$ and $p_{M_j j} = \underset{i}{\max} \{ p_{ij} \}$.
The symmetric measure $\lambda$ likewise takes on values between $0$ and $1$; it equals $0$ when $X$ and $Y$ are independent, and it equals $1$ when each row or column contains at most one nonzero $p_{ij}$.
Moreover, the alternative measure for symmetric situations has been proposed by \cite{kvaalseth2018measuring}.

In this paper, we propose new association measures for asymmetric situations based on multi-categorical proportional reduction in error. 
Naturally, measures applicable in reverse situations, such as $\lambda^{(s)}_{X\vert Y}$, $\lambda^{K(s)}_{X\vert Y}$ for $1 \leq s < r$, can also be considered, allowing for the construction of measures suitable for symmetric situations.
A common method for constructing symmetric measures, as adopted in \cite{doi:10.1080/01621459.1954.10501231}, is the use of a weighted mean.
Additionally, methods such as arithmetic means, geometric means, and harmonic means have been utilized for constructing symmetric measures. 
Although these construction methods are not limited to PRE-based measures, \cite{tomizawa2004generalization} and  \cite{yamamoto2010harmonic} proposed measures based on these approaches.
Furthermore, \cite{urasaki2023generalized} and \cite{yamamoto2010measures} introduced a method employing monotonic functions.
Beyond these, other construction methods include those based on the maximum and minimum values of two asymmetric measures, as exemplified by the well-known symmetric measure, Cram\'er's coefficient \cite{cramer1946mathematical}.

\section{Approximate confidence interval for the measures}\label{sec5}
To evaluate the uncertainty and reliability of our proposed measures in real data analysis, it is essential to construct approximate confidence intervals for these measures.
Let $n_{ij}$ denote the frequency for a cell ($i,j$), and $n=\sum_{i=1}^r\sum_{j=1}^c n_{ij}$ ($i=1,\ldots,r; j=1,\ldots,c$).
Assuming that the observed frequencies $\{ n_{ij}\}$ follow a multinomial distribution, we construct approximate standard errors and large-sample confidence intervals for $\lambda^{(t)}_{Y\vert X}$ and $\lambda^{K(t)}_{Y\vert X}$ using the delta method (e.g. \citealp{bishop2007discrete, agresti2010analysis}).

Letting $\widehat{\lambda}^{(t)}_{Y\vert X}$ denote a plug-in estimator of $\lambda^{(t)}_{Y\vert X}$, the following convergence-in-distribution is applied from the delta method:
\begin{align*}
\sqrt{n} (\widehat{\lambda}^{(t)}_{Y\vert X} - \lambda^{(t)}_{Y\vert X} ) \xrightarrow{d} Normal(0, \sigma^2[\lambda^{(t)}_{Y\vert X}] ),
\end{align*}
so that the estimator $\widehat{\lambda}^{(t)}_{Y\vert X}$ has an approximately normal distribution with mean $\lambda^{(t)}_{Y\vert X}$ and variance $\sigma^2[\lambda^{(t)}_{Y\vert X}]/n$.
The variance is obtained from the partial derivative of $\lambda^{(t)}_{Y\vert X}$ and is given as follows:
\begin{align*}
\sigma^2[\lambda^{(t)}_{Y\vert X}] = \sum_{i=1}^r\sum_{j=1}^c p_{ij}\Delta_{ij}^{2} - \left(\sum_{i=1}^r\sum_{j=1}^c p_{ij}\Delta_{ij}\right)^2 ,
\end{align*}
where 
\begin{align*}
\Delta_{ij} &= \frac{\partial \lambda^{(t)}_{Y\vert X}}{\partial p_{ij}} = \frac{1}{1-\sum^t_{k=1}p_{+ m_{0(k)}}} \left\{1_{A_{(j)}} - \left(1-\lambda^{(t)}_{Y\vert X}\right)1_{B_{(j)}} \right\}.
\end{align*}
Here, the indicator functions $1_{A_{(j)}}$ and $1_{B_{(j)}}$ are defined as:
\begin{align*}
1_{A_{(j)}}=\left\{\begin{matrix}
    1\quad(j\in A)\\
    0\quad(j\notin A)
\end{matrix}\right., \hspace{0.5cm}
1_{B_{(j)}}=\left\{\begin{matrix}
    1\quad(j\in B)\\
    0\quad(j\notin B)
\end{matrix}\right. ,
\end{align*}
with the sets $A = \{m_{i(k)}\vert k=1, \dots, t \}$ and $B = \{m_{0(k)}\vert k=1, \dots, t\}$.
Since $\widehat{\sigma} [\lambda^{(t)}_{Y\vert X} ]$ is a consistent estimator of $\sigma [\lambda^{(t)}_{Y\vert X}]$, $\widehat{\sigma} [\lambda^{(t)}_{Y\vert X}] / \sqrt{n}$ is an estimated standard error for $\widehat{\lambda}^{(t)}_{Y\vert X}$, and $\widehat{\lambda}^{(t)}_{Y\vert X} \pm z_{\alpha/2} \widehat{\sigma} [\lambda^{(t)}_{Y\vert X}] / \sqrt{n}$ is an approximate $100(1-\alpha)\%$ confidence limit for $\lambda^{(t)}_{Y\vert X}$, where $z_{\alpha/2}$ is the upper one-sided normal distribution percentile at level $\alpha/2$.

Similarly, for $\lambda^{K(t)}_{Y\vert X}$, the approximate confidence interval can be constructed by replacing $\Delta_{ij}$ with the following expression:
\begin{align*}
\Delta^{K(t)}_{ij} &=  \frac{\partial \lambda^{K(t)}_{Y\vert X}}{\partial p_{ij}} = \frac{1}{1-\sum^t_{k=1}p_{+ m_{0(k)}}}\left\{\frac{b_i}{a}\left(1_{A_{(j)}} - \frac{b_i}{2} \right) - \left(1 - \lambda^{K(t)}_{Y\vert X}\right)1_{B_{(j)}} \right\},
\end{align*}
where
\begin{align*}
a &= \sqrt{\sum^r_{i=1} \left(\sum^t_{k=1}p_{i m_{i(k)}} \right)^2 /p_{i+}} \;, \quad 
b_i = \sum^t_{k=1}p_{im_{i(k)}} / p_{i+}.
\end{align*}

\section{Numerical experiments}\label{sec6}
In this section, we conduct numerical experiments to examine the behavior and usefulness of the proposed measures. 
The artificial data analysis shows that the proposed association measures behave monotonically as the degree of association increases, while the real data analysis illustrates that the proposed measures can detect associations that are not adequately captured by existing PRE-based measures.

\subsection{Artificial data}
To examine whether the proposed measures $\lambda^{(t)}_{Y\vert X}$ and $\lambda^{K(t)}_{Y\vert X}$ appropriately represent the overall degree of association in contingency tables, we investigate their relationship with the correlation coefficient $\rho$.
For a $2\times 2$ probability table with equal row and column marginal probabilities, $\lambda^{(t)}_{Y\vert X}$ and $\lambda^{K(t)}_{Y\vert X}$ coincide with $\lambda_{Y\vert X}$ and $\lambda^K_{Y\vert X}$, respectively. 
Therefore, their validity in the $2\times 2$ case follows directly from that of the corresponding existing measures.
Since practical contingency tables often have more than two categories, it is important to examine whether the proposed measures still provide reasonable values in multi-categorical settings.
To this end, we constructed $r\times r$ probability tables by discretizing a bivariate standard normal distribution with correlation coefficient $\rho$ into equally spaced intervals. 
Table \ref{sim2} presents examples of $4\times 4$ probability tables obtained in this way for $\rho=0$, $0.4$, and $1$. 
This construction enables us to examine how the values of the proposed measures change as $\rho$ varies, and whether they appropriately reflect the degree of association in probability tables with more than two categories.

\begin{table}[htpb]
\small\centering
\caption{Examples of $4\times 4$ probability tables obtained from a bivariate standard normal distribution with $\rho = 0$, $0.4$, and $1$}
\label{sim2}
\begin{tabular}{ccccc|c}
\hline
 & (1) & (2) & (3) & (4) & Total \\ \hline
& & \multicolumn{2}{c}{$\rho = 0$ } & & \\ 
(1) & 0.0625 & 0.0625 & 0.0625 & 0.0625 & 0.25 \\
(2) & 0.0625 & 0.0625 & 0.0625 & 0.0625 & 0.25 \\
(3) & 0.0625 & 0.0625 & 0.0625 & 0.0625 & 0.25 \\
(4) & 0.0625 & 0.0625 & 0.0625 & 0.0625 & 0.25 \\ \hline
Total & 0.25 & 0.25 & 0.25  & 0.25 & 1 \\ \hline
& &\multicolumn{2}{c}{$\rho = 0.4$ } & \\
(1) & 0.1072 & 0.0692 & 0.0477 & 0.0258 & 0.25 \\
(2) & 0.0692 & 0.0698 & 0.0632 & 0.0477 & 0.25 \\
(3) & 0.0477 & 0.0632 & 0.0698 & 0.0692 & 0.25 \\
(4) & 0.0258 & 0.0477 & 0.0692 & 0.1072 & 0.25  \\ \hline
Total & 0.25 & 0.25 & 0.25 & 0.25 & 1 \\ \hline
& & \multicolumn{2}{c}{$\rho = 1$ } &  \\
(1) & 0.25 & 0 & 0 & 0 & 0.25 \\
(2) & 0 & 0.25 & 0 & 0 & 0.25 \\
(3) & 0 & 0 & 0.25 & 0 & 0.25 \\
(4) & 0 & 0 & 0 & 0.25 & 0.25 \\ \hline
Total & 0.25 & 0.25 & 0.25 & 0.25 & 1 \\ \hline
\end{tabular}
\end{table}

\begin{figure}[htbp]
\centering
\caption{Relationship between $\rho$ and two types of PRE measures for $4\times 4$ and $5\times 5$ probability tables}
\includegraphics[scale=0.22]{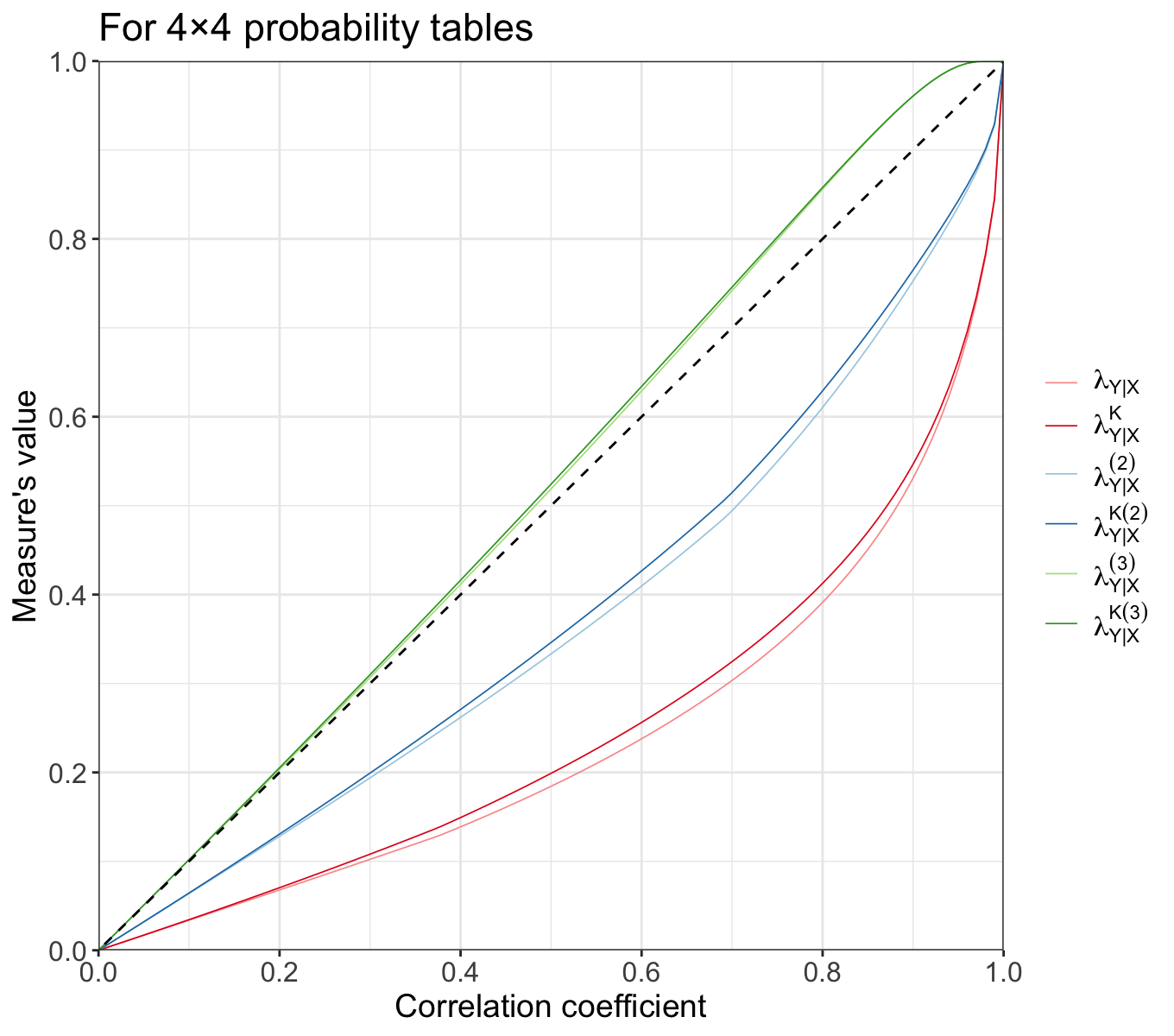}
\includegraphics[scale=0.22]{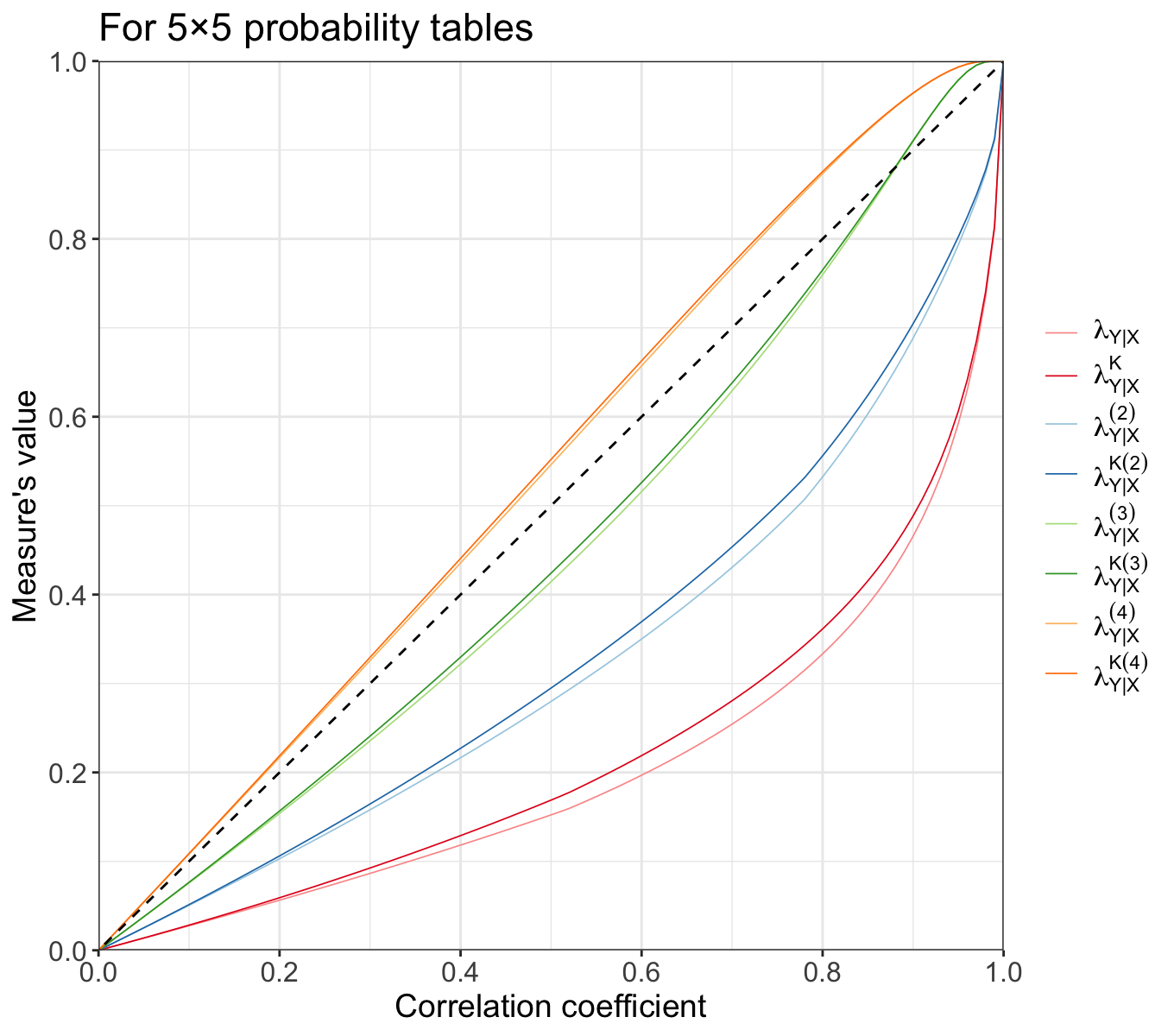}
\label{p}
\end{figure}

Figure \ref{p} shows the relationship between the correlation coefficient $\rho$ and the proposed measures for $4\times 4$ and $5\times 5$ probability tables. 
In each panel, the black dashed line represents the line on which the measure values coincide with $\rho$. 
In the $4\times 4$ case, the red, blue, and green lines correspond to $t=1,2$, and $3$, respectively. 
In the $5\times 5$ case, the red, blue, green, and orange lines correspond to $t=1,2,3$, and $4$, respectively. 
For each value of $t$, the curves of $\lambda^{(t)}_{Y\vert X}$ and $\lambda^{K(t)}_{Y\vert X}$ are very similar.
Figure \ref{p} indicates that all proposed measures increase monotonically with $\rho$ in both the $4\times 4$ and $5\times 5$ cases. 
Thus, the measures consistently reflect the ordering of association strength induced by $\rho$. 
The choice of $t$ mainly affects the rate of increase: smaller values of $t$ yield more moderate changes, whereas larger values of $t$ produce a stronger response to increases in $\rho$. 
This tendency becomes more evident as the number of categories increases.
Although measures with larger $t$ are sometimes numerically closer to $\rho$ in these examples, such closeness is not the primary point of the analysis. 
Rather, the important observation is that the proposed measures behave consistently as the association becomes stronger.
Hence, the results support the usefulness of the proposed measures for evaluating the overall degree of association in multi-category contingency tables.

\subsection{Real data}
As a real data example, consider Table \ref{data}, which presents the cross-classification of children's mental health status (MHS) and their parents' socioeconomic status (SES), based on data from the Midtown Manhattan Study. 
This table has been widely used in the literature on categorical data analysis as a classic example of an ordinal contingency table, and it is available in R through \texttt{data(mentalHealth)} in the \texttt{gnm} package. 
Pearson's chi-squared test for independence yields a p-value less than 0.001, indicating that the null hypothesis of independence is rejected and that an association exists between the two variables. 
In this table, the frequencies are concentrated mainly in the ``Mild'' category of MHS. 
As a result, it may be difficult to evaluate the degree of association appropriately using existing PRE measures.
In this situation, when quantifying the extent to which parental SES is related to children's MHS, it is informative to use our proposed measures for the reasons discussed previously. 
Therefore, using the data, we compare our proposed measures with existing ones, demonstrating their utility.

\begin{table}[!ht]
\footnotesize\centering
\caption{Children's mental health status (MHS) and their parents' socioeconomic status (SES), based on data from the Midtown Manhattan Study}
\label{data}
\begin{tabular}{cccccc}
\hline
 & \multicolumn{4}{c}{Children's MHS} & \\
SES & Well & Mild & Moderate & Impaired & Total \\ \hline
A & 64 & 94 & 58 & 46 & 262 \\
B & 57 & 94 & 54 & 40 & 245 \\
C & 57 & 105 & 65 & 60 & 287 \\
D & 72 & 141 & 77 & 94 & 384 \\ 
E & 36 & 97 & 54 & 78 & 265 \\ 
F & 21 & 71 & 54 & 71 & 217 \\ \hline
Total & 307 & 602 & 362 & 389 & 1660 \\ \hline
\end{tabular}
\end{table}

Table \ref{result} summarizes the analysis of Table \ref{data} using $\lambda^{(t)}_{Y\vert X}$ and $\lambda^{K(t)}_{Y\vert X}$ for $t=1,2$ and $3$.
This table includes the estimated values, standard errors, and $95\%$ confidence intervals for each measure.
The first point of interest is the results for $t=1$.
For Goodman–Kruskal's lambda $\lambda_{Y\vert X}$ ($= \lambda^{(1)}_{Y\vert X}$), the limitation of the measure is evident, as its estimated value is $0$, indicating no association even though Pearson's chi-squared test rejected the null hypothesis of independence and suggested that an association exists in Table \ref{data}.
That is, although the data exhibit an association structure, $\lambda_{Y\vert X}$ fails to capture it at all in this case.
Similarly, the alternative Goodman–Kruskal's lambda $\lambda^K_{Y\vert X}$ ($= \lambda^{K(1)}_{Y\vert X}$) yields an estimated value of only $0.0005$, with a $95\%$ confidence interval of $(-0.0143, 0.0154)$.
Since this interval includes $0$, the result is not consistent with the rejection of independence based on Pearson's chi-squared test.
These results indicate that, when $t=1$, the existing PRE-based measures do not adequately reflect the association structure in Table \ref{data}.

Next, we turn our attention to the results for $t=2$ and $t=3$.
For $\lambda^{(2)}_{Y\vert X}$ and $\lambda^{(3)}_{Y\vert X}$, the estimated values are $0.0598$ and $0.1140$, respectively.
These values suggest that, after excluding the influence of the most dominant categories in the column variable, information on parental SES improves the prediction of children's MHS by $5.98\%$ and $11.40\%$, respectively, compared to when such information is absent.
Moreover, the corresponding confidence intervals, $(0.0080, 0.1120)$ and $(0.0280, 0.2000)$, do not include $0$, supporting the presence of a meaningful association under these settings.
Likewise, $\lambda^{K(2)}_{Y\vert X}$ and $\lambda^{K(3)}_{Y\vert X}$ take values of $0.0610$ and $0.1175$, respectively, with confidence intervals $(0.0100, 0.1120)$ and $(0.0315, 0.2035)$.
These results are broadly similar to those of $\lambda^{(t)}_{Y\vert X}$ and likewise indicate a non-negligible association between parental SES and children's MHS after taking into account the influence of the concentrated frequencies in the ``Mild'' category.

Based on these results, it is demonstrated that our proposed measures possess practical utility in applied contexts.
In particular, unlike the existing PRE-based measures at $t=1$, the proposed measures for $t=2$ and $3$ are able to capture the association structure in the data in a way that is consistent with the rejection of independence.
We believe that both $\lambda^{(t)}_{Y\vert X}$ and $\lambda^{K(t)}_{Y\vert X}$ should be employed in analyses to enable a more comprehensive and multifaceted evaluation of the data.

\begin{table}[!ht]
\small\centering
\caption{Analysis results of Table \ref{data}.}
\label{result}
\begin{tabular}{ccccc}
\hline
Proposed &  & Estimated & Standard  & Confidence \\
measure & $t$ & value & error  & interval \\ \hline
\multirow{3}{*}{$\lambda^{(t)}_{Y\vert X}$} & 1 & $0$ & $0$ & ($0, 0$) \\
& 2 & $0.0598$ & $0.0264$ & ($0.0080$, $0.1120$) \\
& 3 & $0.1140$ & $0.0441$ & ($0.0280$, $0.2000$) \\ \hline
\multirow{3}{*}{$\lambda^{K(t)}_{Y\vert X}$} & 1 & $0.0005$ & $0.0076$ & ($-0.0143$, $0.0154$) \\
& 2 & $0.0610$ & $0.0260$ & ($0.0100$, $0.1120$) \\
& 3 & $0.1175$ & $0.0439$ & ($0.0315$, $0.2035$) \\ \hline
\end{tabular}
\end{table}

\section{Conclusion}\label{sec7}
In this study, we proposed a new class of association measures based on proportional reduction in error. 
The proposed measures extend existing PRE-based measures by incorporating information from multiple categories, rather than relying only on the single most probable category.
In this way, they address a well-known limitation of existing PRE measures, including Goodman–Kruskal's lambda, namely that existing PRE measures may fail to adequately reflect association because they do not fully utilize the information contained in the contingency table.
Like Goodman–Kruskal's lambda, the proposed measures retain a simple and intuitive interpretation in terms of proportional reduction in prediction error. At the same time, by introducing the parameter $t$, they provide a flexible framework for evaluating association across multiple categories. 
This makes it possible to examine the degree of association not only for the most dominant category of the response variable but also for additional categories with relatively large probabilities.
Through numerical experiments, we showed that the proposed measures behave consistently as the degree of association becomes stronger, supporting their usefulness for evaluating the overall degree of association in multi-category contingency tables. 
In the real data analysis, we also demonstrated their practical utility in a setting where the response frequencies were heavily concentrated in a single category. 
In such a case, the existing PRE-based measures with $t=1$ did not adequately capture the association structure, whereas the proposed measures with $t \geq 2$ were able to reveal a meaningful association by taking into account information beyond the most dominant category.
Therefore, the proposed measures $\lambda^{(t)}_{Y\mid X}$ and $\lambda^{K(t)}_{Y\mid X}$ provide useful and interpretable tools for applied research. 
They are particularly valuable when researchers wish to assess association flexibly according to the characteristics of the data, regardless of whether the information relevant to prediction is concentrated in a single category or distributed across multiple categories.

\section*{Acknowledgement}
This work was supported by JST SPRING, Grant Number JPMJSP2151.
Additionally, this work was supported by JSPS Grant-in-Aid for Scientific Research (C) Number JP20K03756.

\appendix
\section{Proof of Theorem 1}\label{secA1}
\begin{proof}[Theorem 1-1]
From $1 \leq t < c $, since $p_{i+}\geq \sum^t_{k=1}p_{i m_{i(k)}} $ holds,
\begin{align*}
\lambda^{(t)}_{Y\vert X} &= \frac{\sum^r_{i=1} \sum^t_{k=1}p_{i m_{i(k)}} - \sum^t_{k=1}p_{+ m_{0(k)}} }{1-\sum^t_{k=1}p_{+ m_{0(k)}}} \\
&\leq \frac{\sum^r_{i=1} p_{i+} - \sum^t_{k=1}p_{+ m_{0(k)}} }{1-\sum^t_{k=1}p_{+ m_{0(k)}}} \\
&= \frac{1 - \sum^t_{k=1}p_{+ m_{0(k)}} }{1-\sum^t_{k=1}p_{+ m_{0(k)}}} \\
&= 1.
\end{align*}
Furthermore, since $\sum^r_{i=1} \sum^t_{k=1}p_{i m_{i(k)}} \geq \sum^t_{k=1}p_{+ m_{0(k)}}$ clearly holds from the definition of $p_{i m_{i(k)}}$ and $p_{+ m_{0(k)}}$,
\begin{align*}
\lambda^{(t)}_{Y\vert X} &\geq 0.
\end{align*}
Hence, $0 \leq \lambda^{(t)}_{Y\vert X} \leq 1$ is proved.
\end{proof}

\begin{proof}[Theorem 1-2]
If $p_{i m_{i(k)}} = p_{i+}p_{+ m_{0(k)}}$ ($1 \leq k \leq t $), then
\begin{align*}
\lambda^{(t)}_{Y\vert X} &= \frac{\sum^r_{i=1} \sum^t_{k=1}p_{i m_{i(k)}} - \sum^t_{k=1}p_{+ m_{0(k)}} }{1-\sum^t_{k=1}p_{+ m_{0(k)}}} \\
&= \frac{\sum^r_{i=1} \sum^t_{k=1}p_{i+}p_{+ m_{0(k)}} - \sum^t_{k=1}p_{+ m_{0(k)}} }{1-\sum^t_{k=1}p_{+ m_{0(k)}}} \\
&= \frac{\sum^r_{i=1} p_{i+} \sum^t_{k=1}p_{+ m_{0(k)}} - \sum^t_{k=1}p_{+ m_{0(k)}} }{1-\sum^t_{k=1}p_{+ m_{0(k)}}} \\
&= \frac{\sum^t_{k=1}p_{+ m_{0(k)}} - \sum^t_{k=1}p_{+ m_{0(k)}} }{1-\sum^t_{k=1}p_{+ m_{0(k)}}} \\
&= 0.
\end{align*}
However, the converse does not hold.
\end{proof}

\begin{proof}[Theorem 1-3]
If $p_{i+} = \sum^t_{k=1} p_{i m_{i(k)}}$ for $i=1,\dots, r$, then
\begin{align*}
\lambda^{(t)}_{Y\vert X} &= \frac{\sum^r_{i=1} p_{i+}  - \sum^t_{k=1}p_{+ m_{0(k)}} }{1-\sum^t_{k=1}p_{+ m_{0(k)}}} \\
&= 1.
\end{align*}
Conversely, when $\lambda^{(t)}_{Y\vert X}=1$,
\begin{align*}
\lambda^{(t)}_{Y\vert X}=1 
&\iff \sum^r_{i=1} \sum^t_{k=1}p_{i m_{i(k)}}  = 1 \\
&\iff \sum^r_{i=1} \left(p_{i+} - \sum^t_{k=1}p_{i m_{i(k)}} \right) = 0 
\end{align*}
Here, since $p_{i+} \geq \sum^t_{k=1}p_{i m_{i(k)}}$, the equality holds when $p_{i+}=\sum^t_{k=1}p_{i m_{i(k)}}$.
Thus, Theorem 1-3 is proved.
\end{proof}

\section{Proof of Theorem 2}\label{secA2}
\begin{proof}[Theorem 2-1]
From $1 \leq t < c $, since $p_{i+}\geq \sum^t_{k=1}p_{i m_{i(k)}} $ holds,
\begin{align*}
\lambda^{K(t)}_{Y\vert X} &= \frac{\sqrt{\sum^r_{i=1} \left(\sum^t_{k=1}p_{i m_{i(k)}} \right)^2 /p_{i+} } - \sum^t_{k=1}p_{+ m_{0(k)}} }{1-\sum^t_{k=1}p_{+ m_{0(k)}}} \\
&\leq \frac{\sqrt{\sum^r_{i=1} p_{i+}^2 /p_{i+} } - \sum^t_{k=1}p_{+ m_{0(k)}} }{1-\sum^t_{k=1}p_{+ m_{0(k)}}} \\
&= 1.
\end{align*}
Furthermore, from Jensen's inequality, we have
\begin{align*}
\sum^r_{i=1} \left(\sum^t_{k=1}p_{i m_{i(k)}} \right)^2 /p_{i+} &= \sum^r_{i=1} p_{i+} \left(\sum^t_{k=1}p_{i m_{i(k)}} /p_{i+} \right)^2 \\
&\geq \left( \sum^r_{i=1} p_{i+} \sum^t_{k=1}p_{i m_{i(k)}} /p_{i+} \right)^2  \\
&= \left( \sum^r_{i=1} \sum^t_{k=1}p_{i m_{i(k)}} \right)^2.
\end{align*}
Therefore, we can obtain
\begin{align*}
\lambda^{K(t)}_{Y\vert X} &\geq \frac{\sqrt{\left( \sum^r_{i=1} \sum^t_{k=1}p_{i m_{i(k)}} \right)^2} - \sum^t_{k=1}p_{+ m_{0(k)}} }{1-\sum^t_{k=1}p_{+ m_{0(k)}}} \\
&= \frac{\sum^r_{i=1} \sum^t_{k=1}p_{i m_{i(k)}} - \sum^t_{k=1}p_{+ m_{0(k)}} }{1-\sum^t_{k=1}p_{+ m_{0(k)}}}.
\end{align*}
Here, since $\sum^r_{i=1} \sum^t_{k=1}p_{i m_{i(k)}} \geq \sum^t_{k=1}p_{+ m_{0(k)}}$ clearly holds from the definition of $p_{i m_{i(k)}}$ and $p_{+ m_{0(k)}}$,
\begin{align*}
\lambda^{K(t)}_{Y\vert X} &\geq 0.
\end{align*}
Hence, $0 \leq \lambda^{K(t)}_{Y\vert X} \leq 1$ is proved.
\end{proof}

\begin{proof}[Theorem 2-2]
If $p_{i m_{i(k)}} = p_{i+}p_{+ m_{0(k)}}$ ($1 \leq k \leq t $), then
\begin{align*}
\lambda^{K(t)}_{Y\vert X} &= \frac{\sqrt{\sum^r_{i=1} \left(\sum^t_{k=1}p_{i+}p_{+ m_{0(k)}} \right)^2 /p_{i+} } - \sum^t_{k=1}p_{+ m_{0(k)}} }{1-\sum^t_{k=1}p_{+ m_{0(k)}}} \\
&= \frac{\sqrt{\sum^r_{i=1}p_{i+} \left(\sum^t_{k=1}p_{+ m_{0(k)}} \right)^2 } - \sum^t_{k=1}p_{+ m_{0(k)}} }{1-\sum^t_{k=1}p_{+ m_{0(k)}}} \\
&= 0.
\end{align*}
Conversely, when $\lambda^{K(t)}_{Y\vert X}=0$, we show by mathematical induction that $p_{i m_{i(k)}} = p_{i+}p_{+ m_{0(k)}}$ ($1 \leq k \leq t $).
\begin{itemize}
\item[(i)] When $t = 1$, $\lambda^{K(1)}_{Y\vert X}=\lambda^{K}_{Y\vert X}$ is obtained, so that $p_{i m_{i(1)}} = p_{i+}p_{+ m_{0(1)}}$ follows from \cite{kvaalseth2018measuring}.
\item[(ii)] When $t = l-1$ (but $l \leq c-1$ from the range of possible values of $t$), we assume that $p_{i m_{i(k)}} = p_{i+}p_{+ m_{0(k)}}$ ($1 \leq k \leq l-1 $) holds for $\lambda^{K(l-1)}_{Y\vert X} = 0$.
Then, when $t = l$, for $\lambda^{K(l)}_{Y\vert X} = 0$,
\begin{align*}
\lambda^{K(l)}_{Y\vert X} = 0 
&\iff \sum^r_{i=1} \left(\sum^l_{k=1}p_{i m_{i(k)}} \right)^2 /p_{i+} = \left( \sum^l_{k=1}p_{+ m_{0(k)}} \right)^2 \\
&\iff \sum^r_{i=1} \left(\sum^l_{k=1}p_{i m_{i(k)}} \right)^2 /p_{i+} = \sum_{i=1}^r p_{i+} \left( \sum^l_{k=1}p_{+ m_{0(k)}} \right)^2 \\
&\iff \sum^r_{i=1} \left( \left(\sum^l_{k=1}p_{i m_{i(k)}} \right)^2 - \left( p_{i+} \sum^l_{k=1}p_{+ m_{0(k)}} \right)^2 \right) /p_{i+} = 0.
\end{align*}
For the above equation, since the assumption holds that $p_{i m_{i(k)}} = p_{i+}p_{+ m_{0(k)}}$ ($1 \leq k \leq l-1 $), we obtain
\begin{align*}
&\left(\sum^l_{k=1}p_{i m_{i(k)}} \right)^2 - \left( p_{i+} \sum^l_{k=1}p_{+ m_{0(k)}} \right)^2 \\
&= \left(\sum^l_{k=1}p_{i m_{i(k)}} + p_{i+} \sum^l_{k=1}p_{+ m_{0(k)}} \right) \left(\sum^l_{k=1}p_{i m_{i(k)}} - p_{i+} \sum^l_{k=1}p_{+ m_{0(k)}} \right) \\
&= \left(\sum^l_{k=1}p_{i m_{i(k)}} + p_{i+} \sum^l_{k=1}p_{+ m_{0(k)}} \right) \left(p_{i m_{i(l)}} - p_{i+}p_{+ m_{0(l)}} \right).
\end{align*}
Therefore, since $p_{i m_{i(l)}} = p_{i+}p_{+ m_{0(l)}}$, it is shown that when $t = l$, $p_{i m_{i(k)}} = p_{i+}p_{+ m_{0(k)}}$ ($1 \leq k \leq l $) holds for $\lambda^{K(l)}_{Y\vert X} = 0$.
\end{itemize}
\end{proof}

\begin{proof}[Theorem 2-3]
If $p_{i+} = \sum^t_{k=1} p_{i m_{i(k)}}$, then
\begin{align*}
\lambda^{K(t)}_{Y\vert X} &= \frac{\sqrt{\sum^r_{i=1} p_{i+}^2 /p_{i+} } - \sum^t_{k=1}p_{+ m_{0(k)}} }{1-\sum^t_{k=1}p_{+ m_{0(k)}}} \\
&= 1.
\end{align*}
Conversely, when $\lambda^{K(t)}_{Y\vert X}=1$,
\begin{align*}
\lambda^{K(t)}_{Y\vert X}=1 
&\iff \sqrt{\sum^r_{i=1} \left(\sum^t_{k=1}p_{i m_{i(k)}} \right)^2 /p_{i+} } = 1 \\
&\iff \sum^r_{i=1} \left(\sum^t_{k=1}p_{i m_{i(k)}} \right)^2 /p_{i+} = 1 \\
&\iff \sum^r_{i=1} \left(p_{i+}^2 -\left(\sum^t_{k=1}p_{i m_{i(k)}} \right)^2 \right)/p_{i+} = 0 \\
&\iff \sum^r_{i=1} \left(p_{i+}+\sum^t_{k=1}p_{i m_{i(k)}}\right) \left(p_{i+} -\sum^t_{k=1}p_{i m_{i(k)}} \right)/p_{i+} = 0
\end{align*}
Here, since $p_{i+} \geq \sum^t_{k=1}p_{i m_{i(k)}}$, the equality holds when $p_{i+}=\sum^t_{k=1}p_{i m_{i(k)}}$.
Thus, Theorem 2-3 is proved.
\end{proof}


\newpage

\bibliographystyle{apacite}
\bibliography{paper-ref}

@book{agresti2010analysis,
  title={Analysis of {Ordinal} {Categorical} {Data}},
  author={Agresti, Alan},
  volume={656},
  year={2010},
  publisher={John Wiley \& Sons, New Jersey}
}

@book{bishop2007discrete,
  title={Discrete {Multivariate} {Analysis}: {Theory} and {Practice}},
  author={Bishop, Yvonne M and Fienberg, Stephen E and Holland, Paul W},
  year={2007},
  publisher={Springer, New York}
}

@book{cramer1946mathematical,
  title={Mathematical {Methods} of {Statistics}},
  author={Cram{\'e}r, Harald},
  year={1946},
  publisher={Princeton University Press}
}

@book{everitt1992analysis,
  title={The {Analysis} of {Contingency} {Tables}},
  author={Everitt, Brian S},
  year={1992},
  publisher={Chapman \& Hall, London}
}

@article{doi:10.1080/01621459.1954.10501231,
author = {Goodman, Leo A and Kruskal, William H},
title = {Measures of Association for Cross Classifications},
journal = {Journal of the American Statistical Association},
volume = {49},
number = {268},
pages = {732-764},
year  = {1954},
publisher = {Taylor \& Francis}
}

@article{goodman1981association,
  title={Association models and canonical correlation in the analysis of cross-classifications having ordered categories},
  author={Goodman, Leo A},
  journal={Journal of the American Statistical Association},
  volume={76},
  number={374},
  pages={320--334},
  year={1981},
  publisher={Taylor \& Francis}
}

@article{goodman1985analysis,
  title={The analysis of cross-classified data having ordered and/or unordered categories: Association models, correlation models, and asymmetry models for contingency tables with or without missing entries},
  author={Goodman, Leo A},
  journal={The Annals of Statistics},
  volume={13},
  number={1},
  pages={10--69},
  year={1985},
  publisher={JSTOR}
}

@article{rom1992generalized,
  title={A generalized model for the analysis of association in ordinal contingency tables},
  author={Rom, Dror and Sarkar, Sanat K},
  journal={Journal of Statistical Planning and Inference},
  volume={33},
  number={2},
  pages={205--212},
  year={1992},
  publisher={Elsevier}
}

@article{kateri2018phi,
  title={$\phi$-divergence in contingency table analysis},
  author={Kateri, Maria},
  journal={Entropy},
  volume={20},
  number={5},
  pages={324},
  year={2018},
  publisher={MDPI}
}

@article{theil1970estimation,
  title={On the estimation of relationships involving qualitative variables},
  author={Theil, Henri},
  journal={American Journal of Sociology},
  volume={76},
  number={1},
  pages={103--154},
  year={1970},
  publisher={University of Chicago Press}
}

@article{tomizawa2004generalization,
  title={Generalization of {Cramer's} coefficient of association for contingency tables: theory and methods},
  author={Tomizawa, Sadao and Miyamoto, Nobuko and Houya, Hidechika},
  journal={South African Statistical Journal},
  volume={38},
  number={1},
  pages={1--24},
  year={2004},
  publisher={South African Statistical Association (SASA)}
}

@book{reynolds1977analysis,
  title={The {Analysis} of {Cross-Classifications}},
  author={Reynolds, Henry T and Reynolds, HT},
  year={1977},
  publisher={Free Press, New York}
}

@article{kvaalseth2018measuring,
  title={Measuring association between nominal categorical variables: an alternative to the {Goodman--Kruskal} lambda},
  author={Kv{\aa}lseth, Tarald O},
  journal={Journal of Applied Statistics},
  volume={45},
  number={6},
  pages={1118--1132},
  year={2018},
  publisher={Taylor \& Francis}
}

@article{kvaalseth2017alternative,
  title={An alternative measure of ordinal association as a value-validity correction of the {Goodman--Kruskal} gamma},
  author={Kv{\aa}lseth, Tarald O},
  journal={Communications in Statistics-Theory and Methods},
  volume={46},
  number={21},
  pages={10582--10593},
  year={2017},
  publisher={Taylor \& Francis}
}

@article{stavig1979alternatives,
  title={Alternatives to {Goodman} and {Kruskal's} lambda},
  author={Stavig, Gordon R},
  journal={Educational and Psychological Measurement},
  volume={39},
  number={4},
  pages={725--731},
  year={1979},
  publisher={Sage Publications Sage CA: Thousand Oaks, CA}
}

@article{yamamoto2010harmonic,
  title={Harmonic, geometric and arithmetic means type uncertainty measures for two-way contingency tables with nominal categories},
  author={Yamamoto, Kouji and Miyamoto, Nobuko and Tomizawa, Sadao},
  journal={Advances and Applications in Statistics},
  volume={17},
  number={2},
  pages={143--159},
  year={2010}
}

@article{urasaki2023generalized,
  title={Generalized {C}ram{\'e}r’s coefficient via f-divergence for contingency tables},
  author={Urasaki, Wataru and Nakagawa, Tomoyuki and Momozaki, Tomotaka and Tomizawa, Sadao},
  journal={Advances in Data Analysis and Classification},
  volume={18},
  pages={893--910},
  year={2024}
}

@article{yamamoto2010measures,
  title={Measures of proportional reduction in error for two-way contingency tables with nominal categories},
  author={Yamamoto, K and Tomizawa, S},
  journal={Biostatistics, Bioinformatics and Biomathematics},
  volume={2},
  pages={43--52},
  year={2010}
}

@article{tichy2011evaluating,
  title={Evaluating the stability of the classification of community data},
  author={Tich{\`y}, Lubom{\'\i}r and Chytr{\`y}, Milan and Smarda, Petr},
  journal={Ecography},
  volume={34},
  number={5},
  pages={807--813},
  year={2011},
  publisher={Wiley Online Library}
}

@article{jaroszewicz2004goodman,
  title={The {Goodman-Kruskal} coefficient and its applications in genetic diagnosis of cancer},
  author={Jaroszewicz, Szymon and Simovici, Dan A and Kuo, Winston Patrick and Ohno-Machado, Lucila},
  journal={IEEE Transactions on Biomedical Engineering},
  volume={51},
  number={7},
  pages={1095--1102},
  year={2004},
  publisher={IEEE}
}

@article{suich2003asymptotic,
  title={An asymptotic partial correlation test for the {Goodman-Kruskal} $\lambda$},
  author={Suich, Ronald C and Turek, Richard J},
  journal={British Journal of Mathematical and Statistical Psychology},
  volume={56},
  number={1},
  pages={111--117},
  year={2003},
  publisher={Wiley Online Library}
}

@article{costner1965criteria,
  title={Criteria for measures of association},
  author={Costner, Herbert L},
  journal={American Sociological Review},
  volume={30},
  number={3},
  pages={341--353},
  year={1965},
  publisher={JSTOR}
}

@article{kvaalseth2023association,
  title={Association Between Nominal Categorical Variables: New Measure Formulation Based on Metric Distances and Value Validity},
  author={Kv{\aa}lseth, Tarald O},
  journal={Journal of Statistical Theory and Practice},
  volume={17},
  number={4},
  pages={47},
  year={2023},
  publisher={Springer}
}

\end{document}